\documentclass[]{spie} 
\usepackage[]{graphicx}
\usepackage{amsmath}
\usepackage{color}
\newcommand{\ms}{m s$^{-1}$ }

\title{The Habitable-zone Planet Finder Calibration System}

\author{Samuel Halverson\supit{a,b},  Suvrath Mahadevan\supit{a,b}, Lawrence Ramsey\supit{a,b}, Ryan Terrien\supit{a,b}, Arpita Roy\supit{a,b}, Christian Schwab\supit{a}, Chad Bender\supit{a,b}, Fred Hearty\supit{a}, Eric Levi\supit{a}, Steve Osterman\supit{c}, Gabe Ycas\supit{d}, Scott Diddams\supit{d} 
\skiplinehalf
\supit{a}Department of Astronomy \& Astrophysics, The Pennsylvania State University, 525 Davey Laboratory, University Park, 16802, USA; \\
\supit{b}Center for Exoplanets \& Habitable Worlds, The Pennsylvania State University, University Park, PA 16802; \\
\supit{c} Johns Hopkins University Applied Physics Laboratory, Laurel, MD 20723, USA; \\
\supit{d}Time and Frequency Division, National Institute of Standards and Technology, Boulder, CO 80305, USA; \\
}

\authorinfo{E-mail: shalverson@psu.edu}
 
\begin{document} 
  \maketitle 

\begin{abstract}
We present the design concept of the wavelength calibration system for the Habitable-zone Planet Finder instrument (HPF), a precision radial velocity (RV) spectrograph designed to detect terrestrial-mass planets around M-dwarfs. HPF is a stabilized, fiber-fed, R$\sim$50,000 spectrograph operating in the near-infrared (NIR) z/Y/J bands from 0.84 to 1.3 microns. For HPF to achieve 1 \ms or better measurement precision, a unique calibration system, stable to several times better precision, will be needed to accurately remove instrumental effects at an unprecedented level in the NIR. The primary wavelength calibration source is a laser frequency comb (LFC), currently in development at NIST Boulder, discussed separately in these proceedings. The LFC will be supplemented by a stabilized single-mode fiber Fabry-Perot interferometer reference source and Uranium-Neon lamp. The HPF calibration system will combine several other new technologies developed by the Penn State Optical-Infrared instrumentation group to improve RV measurement precision including a dynamic optical coupling system that significantly reduces modal noise effects. Each component has been thoroughly tested in the laboratory and has demonstrated significant performance gains over previous NIR calibration systems.
\end{abstract}


\section{Introduction}
Wavelength calibration is a critical element in any high resolution spectrograph, particularly in the context of precision Doppler measurements. A stable frequency source referenced to an atomic standard is needed to accurately discern instrumental drift from stellar radial velocity shifts. Molecular I$_2$ absorption cells and Thorium-Argon emission lamps have sufficient line density to yield 1 \ms overall instrument precision for stabilized high resolution optical instruments\cite{2013PASP..125.1336T, 2003Msngr.114...20M}, but have few features at near-infrared (NIR) wavelengths. Currently, no comparable atomic lamps or molecular absorption cells have demonstrated similar measurement precisions in the NIR. Future NIR Doppler instruments such as the Habitable-zone Planet Finder (HPF\cite{2012SPIE.8446E..1SM}) on the 10-m Hobby-Eberly Telescope (HET) and the Calar Alto high-Resolution search for M dwarfs with Exo-earths with Near-infrared and optical Echelle Spectrographs (CARMENES\cite{2011IAUS..276..545Q}) on the 3.5 m telescope at the Calar Alto Observatory will require specialized wavelength calibration systems as these instruments aim to achieve 1 \ms precision in the NIR. 

This paper summarizes the overall design of the HPF calibration system and discusses some of the new technologies incorporated to increase calibration precision. Section~\ref{sec:sources} discusses the array of wavelength references available to HPF for calibration. Section~\ref{sec:mn} outlines a new modal noise reduction system for coherent calibration sources. An overview of the optomechanical design of the complete system is presented in Section~\ref{sec:opto}.

\section{Calibration Sources}
\label{sec:sources}
\begin{figure}
\begin{center}
\includegraphics[width=5.5in]{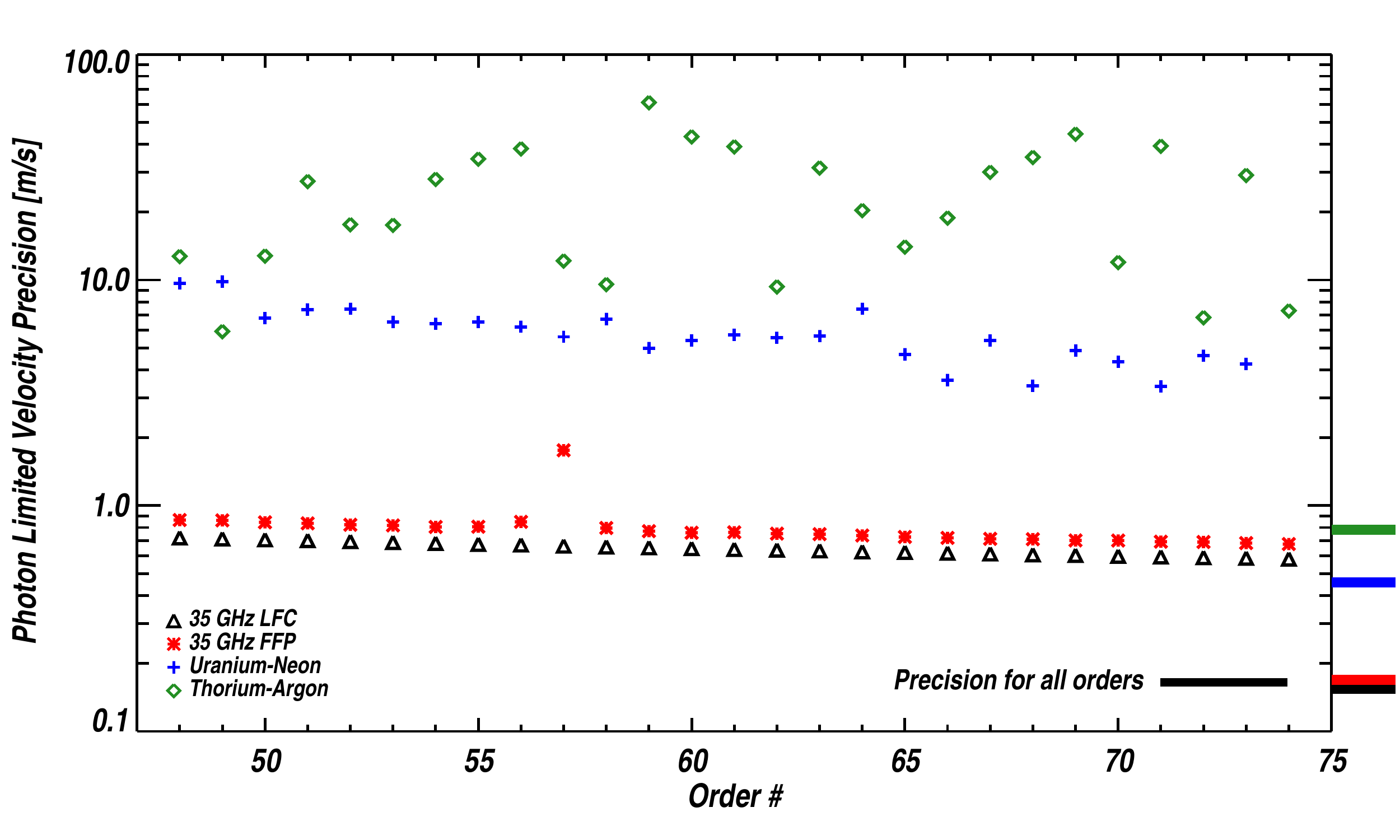}

\caption{Theoretical photon-limited velocity precision\cite{2001A&A...374..733B} comparison of NIR wavelength calibration sources for each order in the HPF bandpass. Simulations assume a resolution of 50,000, three-pixel sampling, and a signal-to-noise of 200 in the extracted spectrum. The LFC and fabry-perot sources give significantly better measurement precisions than atomic lamp sources.}

\label{fig:cal_comp}
\end{center}
\end{figure}

HPF will use a suite of wavelength calibration sources to complete a wide range of science goals. The primary calibration source for high precision radial velocity measurements will be a broadband Laser Frequency Comb (LFC), discussed separately in these proceedings.  The LFC produces a dense grid of emission features, stable at the $\sim$10 c\ms level\cite{2012OExpr..20.6631Y}, across the entire instrument focal plane. The frequency comb has a much richer intrinsic spectrum than a typical atomic lamp (see Figure~\ref{fig:cal_comp}) and is the ideal source for high precision wavelength calibration. A custom, stabilized single-mode fiber (SMF) fabry-perot (FFP) etalon will also be available for precise short-term calibration. Tests with a prototype H-band FFP device showed the device to be stabile to $<1$ \ms\cite{2014PASP..126..445H}. The HPF FFP will build on the success of the prototype device and yield relative higher measurement precision than typical atomic lamps. Uranium-Neon and Thorium-Argon lamps\cite{2010SPIE.7735E.231R} will be used as secondary calibration sources for more coarse wavelength calibration. 

\section{Modal Noise Reduction System}
\label{sec:mn}
Several next generation spectrographs, including HPF, will use coherent emission line calibration sources (like LFCs and Fabry-Perot cavities) to achieve precision sufficient to detect earth-size planets. Many of these photonic sources couple light from single-mode fibers into the larger core multi-mode fibers used in these spectrographs. This coupling leads to very few modes being excited, thereby exacerbating the \textit{modal noise} measured by the spectrograph.  Commercial fiber mode scramblers have been used to reduce this issue for visible wavelength instruments, but are not the complete solution in the NIR due to the decreased number of modes at longer wavelengths ($\mathrm{N_{modes}}\propto\lambda^{-2}$).  This is an increasingly pressing problem for high precision RV spectrographs that require both high stability and high signal-to-noise. As a prime example of modal noise effects: our tests with a NIR LFC on the Pathfinder spectrograph\cite{2010SPIE.7735E.231R} were limited at 10 \ms due to modal noise, despite the LFC being intrinsically stable to significantly higher precision\cite{2012OExpr..20.6631Y}. Bulk mechanical agitation of the fiber can be sufficient for continuum sources (e.g. starlight) but does not suppress modal noise at levels required for high precision RV measurements when using narrow line-width calibration sources.

\begin{figure}
\begin{center}
\includegraphics[width=5in]{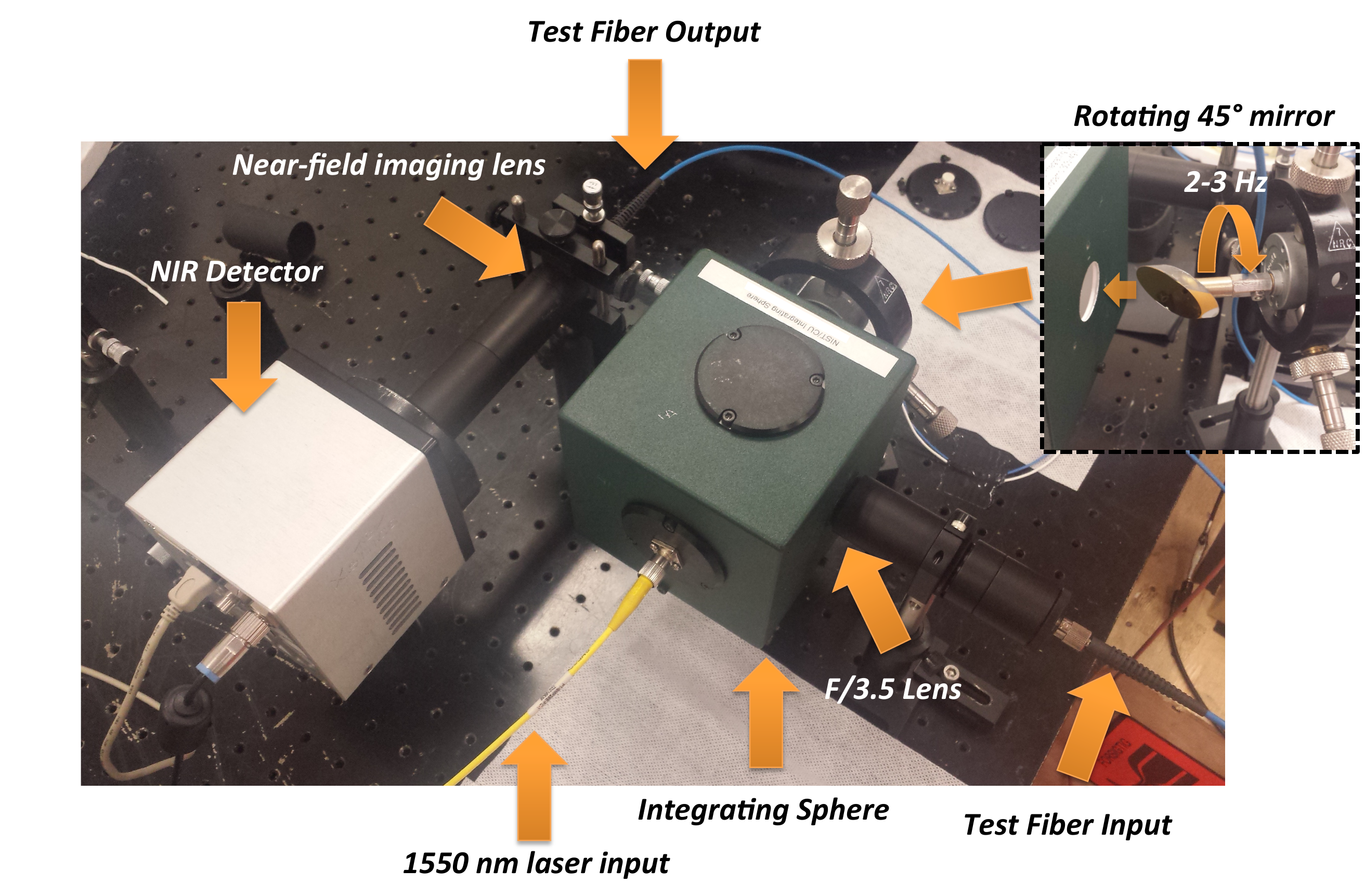}

\caption{Laboratory modal noise measurement apparatus. Light from a narrow-band 1550 nm fiber laser is coupled to a 300 micron test fiber through a NIR optimized integrating sphere. A 45 degree mirror on a motorized rotation stage is placed opposite to the illumination fiber. This ensures the input beam is periodically redirected within the sphere during a given exposure, thereby maximizing modal mixing in the output fiber.}

\label{fig:MN_setup}
\end{center}
\end{figure}

\begin{figure}
\begin{center}
\includegraphics[width=6.7in]{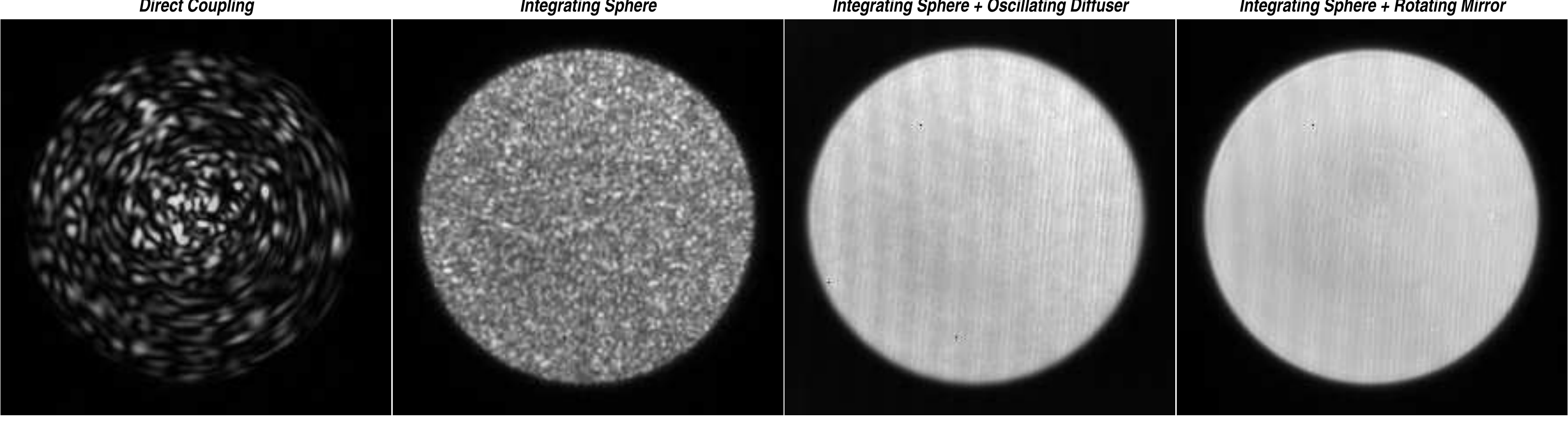}
\includegraphics[width=6.7in]{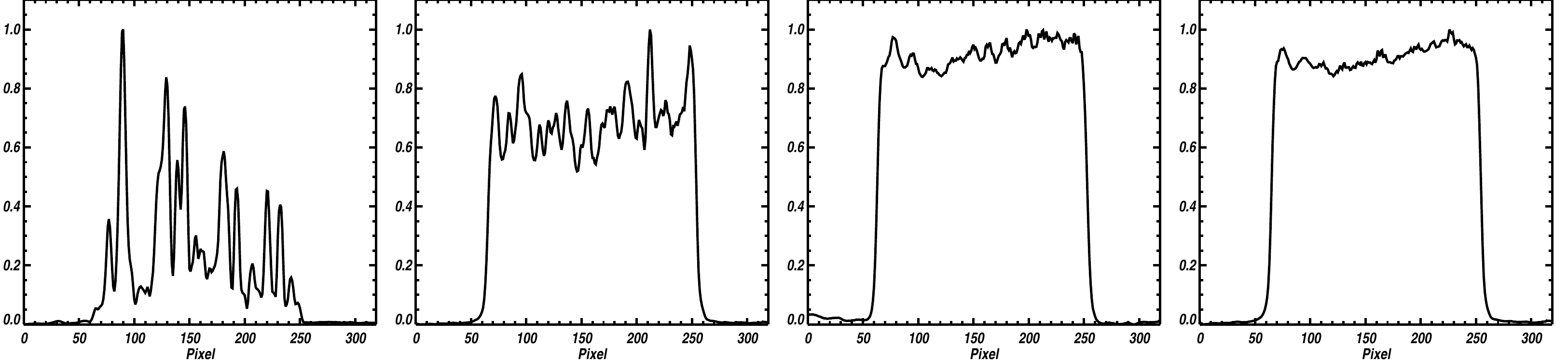}

\caption{Top: Near-field images of 300 micron multi-mode fiber illuminated with a narrow-band (1 MHz line-width) 1550 nm SMF laser. Each image is taken using a different fiber coupling configuration. In all cases the 300 micron test fiber is illuminated with an F/3.5 beam, roughly equivalent to the illumination from the HET fore-optics. The vertical fringe patterns are due to interference at the detector window. Bottom: Intensity profiles of near field images. The combination of integrating sphere and rotating mirror produce the smoothest near-field image with no visible speckles.}

\label{fig:mod_noise_ims}
\end{center}
\end{figure}

Our solution to this issue is to use a \textbf{combination of diffusion and temporal illumination variation} at the fiber input to maximize mode mixing within the fiber. Our laboratory test apparatus is shown in Figure~\ref{fig:MN_setup}. Light from a narrow-band (1 MHz) 1550 nm laser is coupled to a 300 micron test fiber through an integrating sphere. This increases the number of injected modes propagating through the fiber and provides a very diffuse illumination of the fiber face. Combining the integrating sphere with a rotating mirror element (placed opposite to the single-mode fiber illumination port) further suppresses speckle-induced noise in the fiber output and yields a very stable near-field illumination (see Figure~\ref{fig:mod_noise_ims}). The rotating mirror periodically redirects the input beam within the sphere, further increasing the mode mixing in the fiber. This technique has the benefit of not requiring a mechanical agitator that stresses the fiber and can be used with any calibration source, though is restricted to sources that are bright enough (such as frequency combs and fabry-perot etalons) to yield sufficient flux after considering the losses associated with the integrating sphere. We previously tested a commercial oscillating diffuser element which, when combined with the integrating sphere, significantly reduced modal noise effects\cite{2014ApJ...786...18M} but still resulted in some visible speckles at the fiber output. The rotating mirror system results in a smooth output illumination that is insensitive to fiber bends, even when using a monochromatic laser source.

\section{Optomechanical Design}
\label{sec:opto}
\begin{figure}
\begin{center}
\includegraphics[width=5in]{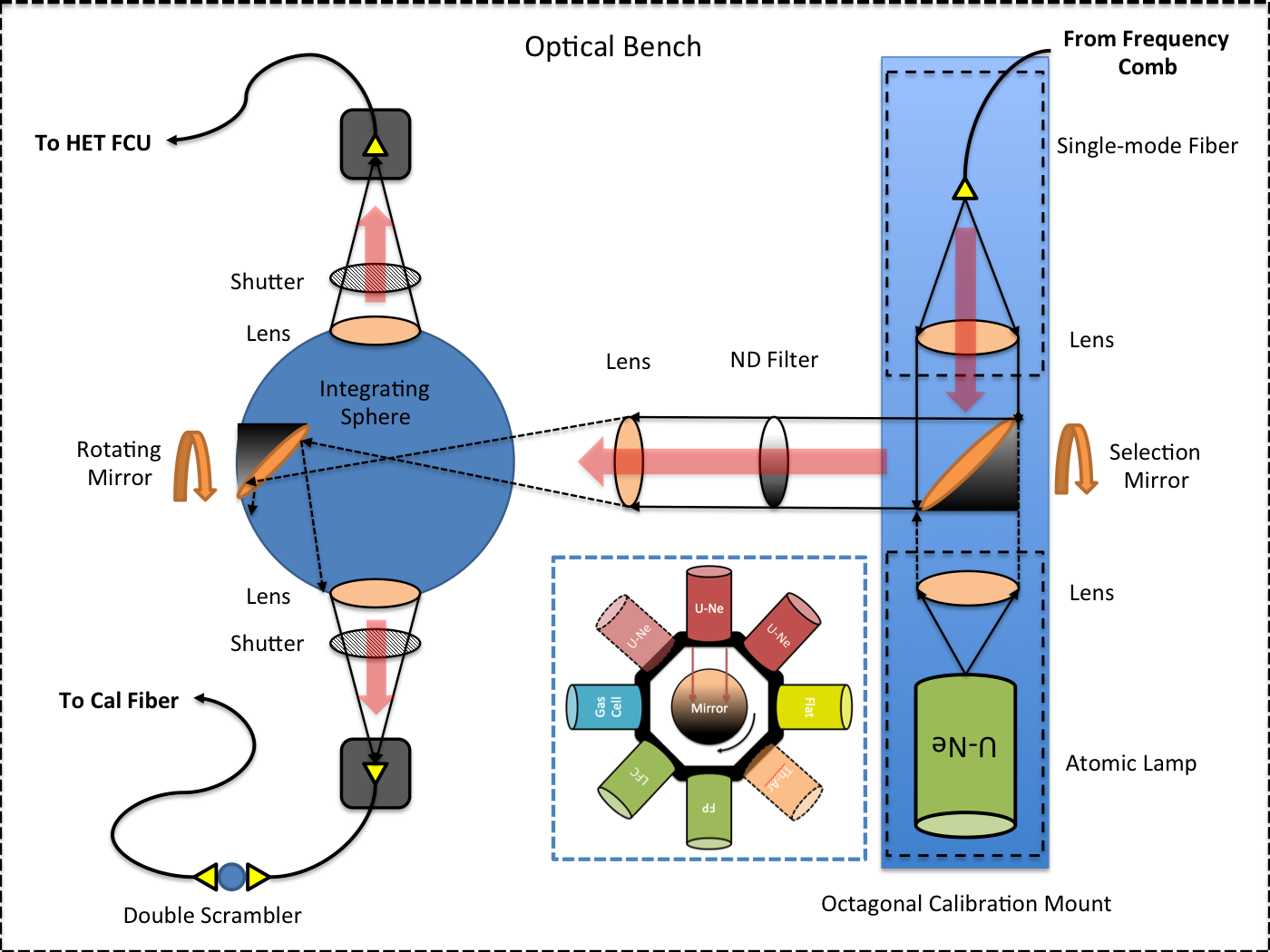}

\caption{Optomechanical design concept of HPF calibration system. Light is coupled to two separate multi-mode fibers through a simple lens system that mimics the input beam from the telescope. An off-axis mirror is inserted into the integrating sphere opposite to the illumination beam and rotated using a DC motor. The mirror scrambles the fiber mode distribution and reduces modal noise in the output. A neutral-density filter wheel set is used to match flux levels of the calibration fiber and science fiber.}

\label{fig:bench_schem}
\end{center}
\end{figure}

The primary purpose of the HPF calibration system is to couple light from a calibration source to the dedicated calibration fiber entering the spectrograph. The system must also reduce modal noise effects and have the ability to couple light from a calibration source into the primary instrument science fiber through the telescope top-end optics (see Figure~\ref{fig:bench_schem}).

\begin{figure}
\begin{center}
\includegraphics[width=3.35in]{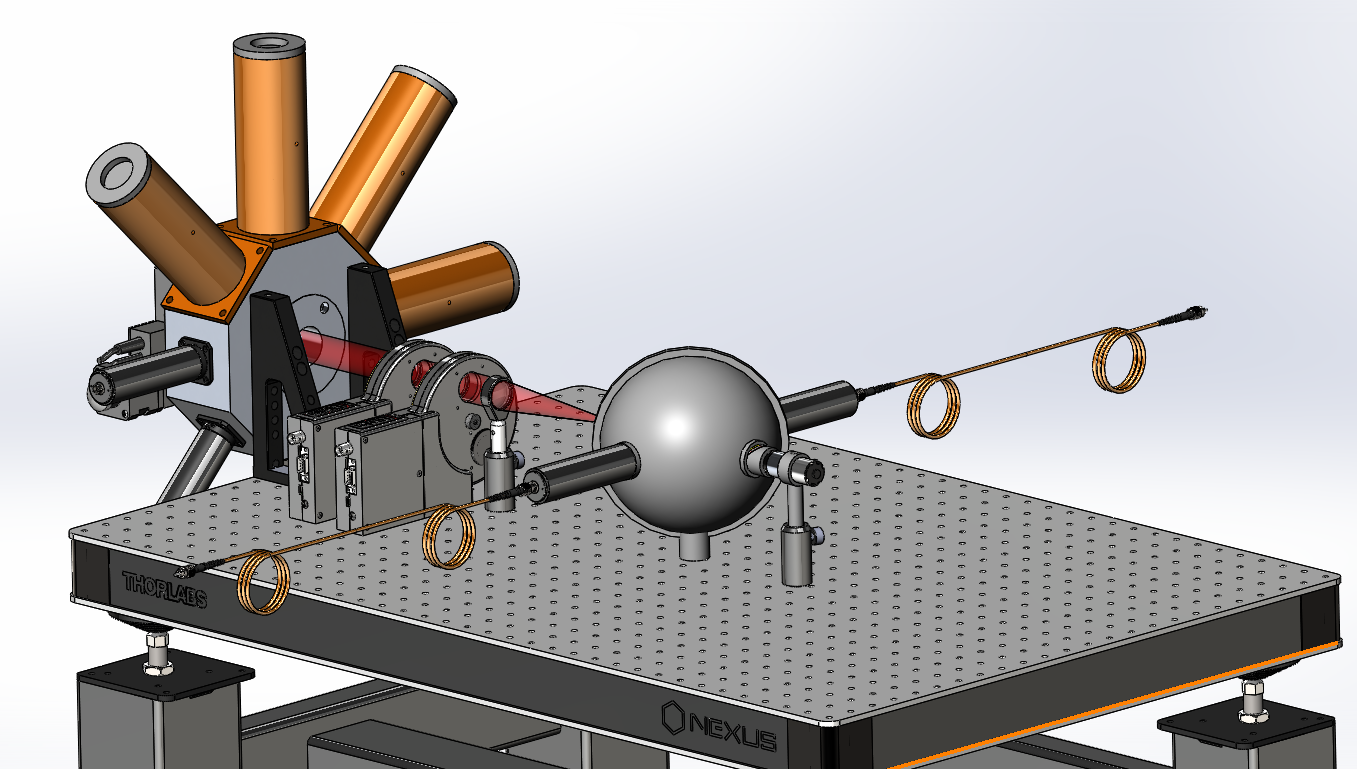}
\includegraphics[width=3.35in]{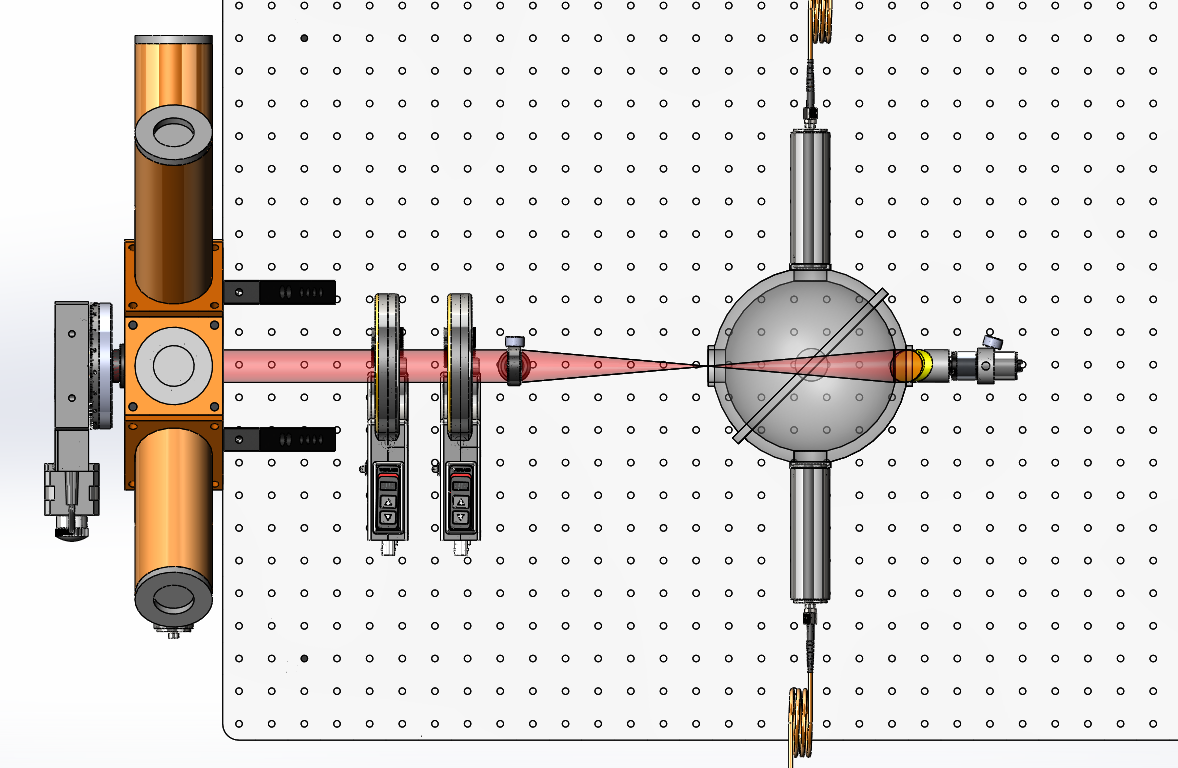}

\caption{3D Renderings of HPF calibration system. All components will mount to a 2' x 3' optical breadboard. The majority of optomechanical parts are commercial off-the-shelf products.}

\label{fig:bench_render}
\end{center}
\end{figure}

The calibration system is built around an octagonal calibration source mount, inspired by the CARMENES calibration unit design\cite{2012SPIE.8446E..0RQ}, that houses a suite of calibration lamps and optical fiber collimators. A rotating mirror selects one of eight calibration slots on a fixed octagonal mount. Light from a given source is then focused into an integrating sphere, reflected off a rotating mirror, and coupled into the primary calibration fiber. A second fiber is used to couple calibration light into the science fiber through the HET Field Calibration Unit (FCU)\cite{2012SPIE.8444E..4JL}. The majority of components are commercially available, off-the-shelf products and will be combined on a single 2\lq{} x 3\lq{} optical bench. Figure~\ref{fig:bench_render} shows the full calibration system model. 

\section{Summary}
We present the baseline design of the calibration system for the Habitable-zone Planet Finder instrument. The calibration system incorporates several new technologies developed specifically to improve the overall measurement precision of the HPF spectrograph. A dynamic optical coupling system that combines a rotating mirror assembly and an integrating sphere will be implemented to mitigate modal noise issues for the planned broadband frequency comb calibration source. We are also developing a fiber fabry-perot interferometer that will be stable to $<$1 \ms to supplement the laser comb\cite{2014PASP..126..445H}. All components have been thoroughly tested in the laboratory and, when combined with the frequency comb, will enable HPF to reach 1 \ms measurement precision. 

\acknowledgments{This work was partially supported by funding from the Center for Exoplanets and Habitable Worlds. The Center for Exoplanets and Habitable Worlds is supported by the Pennsylvania State University, the Eberly College of Science, and the Pennsylvania Space Grant Consortium. We acknowledge support from NSF grants AST 1006676, AST 1126413, AST 1310885, and the NASA Astrobiology Institute (NNA09DA76A) in our pursuit of precision radial velocities in the NIR. SPH acknowledges support from the Penn State Bunton-Waller, and Braddock/Roberts fellowship programs and the Sigma Xi Grant-in-Aid program.}

\bibliography{HPF_Calibration}   
\bibliographystyle{spiebib} 

\end{document}